\documentclass[vecphys]{svmult}

\usepackage{makeidx}         
\usepackage{graphicx}        
\usepackage{multicol}        
\usepackage[bottom]{footmisc}

\makeindex             

\begin{document}

\title*{Molecular absorptions in high-z objects}
\titlerunning{High-z molecular absorptions} 
\author{F. Combes}
\institute{LERMA, Observatoire de Paris, 61 Av. de 
l'Observatoire, F-75014, Paris, France
\texttt{francoise.combes@obspm.fr}}
\maketitle

\begin{abstract}
\vskip  -3mm
Molecular absorption lines measured along the line of sight
of distant quasars are important probes of the
gas evolution in galaxies as a function of redshift.
 A review is made of the handful
of molecular absorbing systems studied so far,
 with the present sensitivity of mm instruments.
They produce information on the chemistry of the ISM
at z $\sim$ 1, the physical state of the gas, in terms of clumpiness,
density and temperature. The CMB temperature can be derived
as a function of z, and also any possible variations of fundamental
constants can be constrained. With the sensitivity of ALMA, many
more absorbing systems can be studied, for which
some predictions and perspectives are described.
\vskip  -6mm
\end{abstract}

\section{Introduction}
\label{sec:1}

Molecular absorptions at intermediate redshift began
to be studied more than a decade ago, after the discovery
of CO absorption in front of the BL Lac object PKS1413+135 at 
z=0.25 (Wiklind \& Combes 1994). Although many groups
undertook active searches, there are still now
 only 5 molecular absorbing systems detected at high z: PKS1413+135 and
B3 1504+377, which are self-absorbing systems, and 3 gravitational lens
systems B0218+357, PKS1830-211, PMN J0134-0931 (with OH only).
Table 1 summarises the properties of these systems, together
with the few local extra-galactic ones.

With respect to emission, absorption measurements are 
quite sensitive, even to a small amount of molecular gas
along the line of sight. The detection  
depends mainly on the background source intensity,
and the rarity of the detections until now is due
to that of strong millimetric radio sources. Due to its
sensitivity increase, there could be
 $\sim$ 30-100 times more sources detected
with ALMA.

{\bf Scientific Goals}

The study of molecular absorbers at
high-z allows to reach several goals:

{\bf 1--}
to detect molecules at high z with
 much more sensitivity (down to 1 M$_\odot$) 
than with emission and with complementary insight

{\bf 2--}
to study the evolution with z of chemical abondances:
 not only CO lines are detectable, but molecular surveys are possible

{\bf 3--}
to measure the CMB temperature as a function of redshift,
to independently estimate the Hubble constant, through the time
delay between two gravitational lens images

{\bf 4--}
to probe the variation of fundamental constants ($\alpha$, g$_p$, $\mu$= m$_e$/m$_p$).
Several theories based on superstrings, Kaluza-Klein theory, or compactified extra-dimensions,
 predict spatio-temporal variations of the fundamental constants
 (Uzan 2003, Murphy et al 2003, Chand et al 2006).

{\bf New local absorptions}

The Centaurus A (NGC 5128) dust lane is well known
to absorb in front of the strong internal radio source, and
the absorption is diluted in the emission for the CO lines
(the same phenomenon is occurring also for M82 in a lesser extent).
The absorption is however completely detached for the high density tracers,
like HCO$^+$ and HCN (Wiklind \& Combes 1997a).
The absorption extends over very broad wings, suggesting
perturbed kinematics, or outflows. 

Recently, very broad absorption extending to the blue wing
was observed in the HI line towards 3C 293 by  Morganti et al (2003): the
total width of 1400km/s  absorption implies neutral gas
entrained in the radio jet towards the observer. This observation
is confirming theoretical expectations of AGN feedback on the 
interstellar medium of the host galaxies. The 3C 293 host galaxy  has
already been observed in the CO line, and found quite rich in molecular gas
(Evans et al 1999).  Garcia-Burillo et al (2006) have observed this strong radio
source at 1mm and 3mm with the IRAM interferometer, and 
found several absorption components, in CO, HCO$^+$ and HCN.
The high resolution helps to disentangle absorption from emission
in the nucleus. The shape of the CO emission map suggests an 
interaction between the jet and the ISM, able to redirect the jet,
and  produce the HI outflow. The molecular lines are however not as broad
as the HI line. The HCO$^+$ absorption has not only a component 
in front of the nucleus, but also in front of the radio jet.
Strong HCO$^+$ and CO absorptions are also detected in front
of 4C 31.04, clearly on the blue-side of the total spectrum, delineated
by emission.

\begin{table}
\centering
\caption{Brief census of molecular absorbers in radio}
\begin{tabular}{lllllll}
\hline\noalign{\smallskip}
Source & z$_a^1$ & z$_e^2$ & N$_c^3$  & N(H$_2$)$^4$  & $\Delta$V$^5$ & Molecules \\
       &       &       &        & cm$^{-2}$ &   km/s    &           \\
\noalign{\smallskip}\hline\noalign{\smallskip}
Cen-A & 0.0018 & 0.0018 &  17   &  2.0 10$^{20}$ & 80.  & CO, HCN, HCO$^+$, N$_2$H$^+$, CS...  \\
3C 293 & 0.045 & 0.045 &  3   &  1.5 10$^{19}$ & 40.  & CO, HCN, HCO$^+$  \\
4C 31.04 & 0.06 & 0.06 &  2   &  1.0 10$^{19}$ & 120.  & CO, HCN, HCO$^+$  \\
PKS1413+135 & 0.247 & 0.247 &  2   &  4.6 10$^{20}$ & 2.  & CO, HCN, HCO$^+$, HNC  \\
B3 1504+377 & 0.673 & 0.673 &  2   &  1.2 10$^{21}$ & 75.  & CO, HCN, HCO$^+$, HNC  \\
B 0218+357 & 0.685 & 0.94 &  1   &  4.0 10$^{23}$ & 20.  & CO, HCN, HCO$^+$, H$_2$O, NH$_3$, H$_2$CO  \\
PMN J0134-0931 & 0.765 & 2.22 &  3   &  -- & 100.  & OH  \\
PKS1830-211 & 0.885 & 2.51 &  2   &  4.0 10$^{22}$ & 40.  & CO, HCN, HCO$^+$, N$_2$H$^+$, CS...   \\
\noalign{\smallskip}\hline
\end{tabular}
$^1$ redshift of absorption lines;
$^2$ redshift of background continuum source;
$^3$ number of components in absorption;
$^4$ maximum H$_2$ column density over components;
$^5$ maximum velocity width 
\end{table}

\vskip  -6mm
\section{Higher redshift absorptions}
\label{sec:2}

After the first system PKS 1413+135, another internal
absorption was detected with several components in
B3 1504+377 (Wiklind \& Combes 1996a).
Then the absorption was detected in intervening systems,
which amplifies the background quasar by lensing effects
(B0218+357, Wiklind \& Combes 1995, Menten et al 1996, Gerin et al 1997);
in front of PKS1830-211, the redshift of the lens was found by 
sweeping the band over 14 GHz. This meant observing with 14 tunings, before
detecting  2 absorption lines and determining  unambiguously the redshift
(Wiklind \& Combes 1996b).  The third gravitationally lensed quasar is
PMN J0134-0931, detected only in the OH lines, but not in CO or
HCO$^+$ (Kanekar et al 2005).
 
The absorbing redshifts range up to z$\sim$1 (the background quasar up to z$\sim$2), 
and it becomes difficult to find higher redshift radio sources, that are strong 
enough in the millimeter domain. The synchroton spectrum
is frequently steep, and the intensity fades at high frequency.
This means that the K-correction plays a very negative role here.
In addition, the number of quasars per comoving volume is expected
to decrease after z=2.
A solution is to follow the high-z quasars and their emission 
in red-shifting our observations, down to the centimeter domain.

With the present instrumentation at IRAM, a full search was undertaken 
with selection of candidates as:
{\bf 1--} strong mm source (at least 0.15 Jy at 3mm), about 150 sources,
searching at the host redshift when known,
{\bf 2--} or at a different z, if already an absorption is detected in HI-21cm, 
or DLAs, or MgII or CaII (e.g. Carilli et al 1993),
{\bf 3--} all mm-strong radio-source with a known gravitational lens (VLBI)
(Webster et al 1995, Stickel \& K\"uhr 1993, Jackson \& Browne 2007).
This survey led to mostly negative results, meaning that a much larger 
sample of radio-sources, and much more sensitivity is required to 
find more molecular absorption systems.

\vskip  -6mm
\subsection{Absorption in the quasar host}

The BL Lac PKS1413+135 at z=0.247
is an edge-on galaxy, and the nucleus is obscured
by  Av $>$ 30 mag (McHardy et al 1994).
On the line of sight, a
very narrow absorption  $<$ 1km/s  has been found:
since the continuum source is 
highly  variable, it was possible to probe the small
scale structure of the interstellar medium 
(Wiklind \& Combes 1997b).

Towards B3 1504+377 at z=0.672,
7 different molecular absorption lines are detected, with 
a large separation 330km/s, which could be explained by
a highly non-circular motions in the center, with
a more regular spiral arm in the outer parts.
The observed HNC/HCN absorption ratio implies thermalization
of the gas, with the excitation temperature equal to 
the kinetic one. As is frequently observed in molecular 
absorptions, the 
HCO$^+$ is enhanced by 10-100, which can only be
explained by a combination of a
diffuse and a clumpy medium  (e.g. Lucas \& Liszt 
1994, 98).

\vskip  -6mm
\subsection{Absorption in the intervening lens galaxy}

{\bf B0218+357}

B0218+357 is amplified by  a
gravitational lens at z=0.685: the source is split
in 2 main images A and B, with an Einstein ring (cf Figure 1). 
In VLBI, the A and B components reveal a
detailed structure, with two bright cores and
extended radio jet components (Biggs et al 2003).
It is the absorber with the largest column density 
around 10$^{24}$ cm$^{-2}$ at maximum.
All three CO isotopic lines up to C$^{18}$O
are optically thick (Combes \& Wiklind 1995).

\begin{figure}
\centering
\includegraphics[width=11cm]{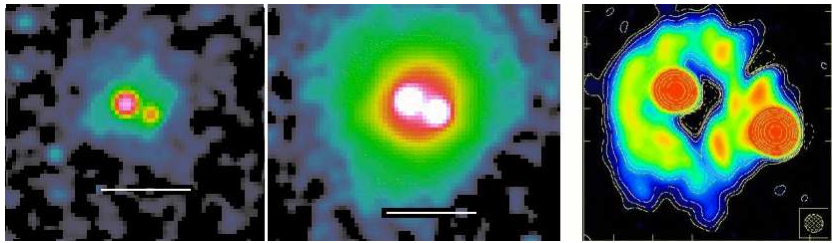}
\caption{The B0218+357 gravitational lens imaged
by the HST in 2 bands, V ({\bf Left})  and H ({\bf Middle} Jackson et al 2000 and the
CASTLES collaboration; in the left and middle panels, the white bar is 1 arcsec), and in radio with 
JIVE ({\bf Right}, Biggs et al 1999). The distance between the two gravitational images is 0.335 arcsec.}
\label{fig:1}
\end{figure}

This has allowed the search of many molecules, and
in particular important ones undetected in our Galaxy
due to atmospheric absorption at z=0.
Search for O$_2$ lines at 56, 119, 368 and 424 GHz in the 
rest frame have led to upper limits O$_2$/CO $<$ 2 10$^{-3}$ 
(Combes \& Wiklind 1995,  Combes et al 1997), suggesting that 
most of the oxygen should be in the form of OI.
The H$_2$O molecule at 557 GHz has been detected, and tentatively
LiH at 444GHz in the rest frame (Combes \& Wiklind 1997, 98),
with H$_2$O/H$_2$=10$^{-5}$ and LiH/H$_2$ $\sim$ 3 10$^{-12}$.
NH$_3$ has been detected at 2cm (Henkel et al 2005),

Recent deep HST imagery reveals the lensing
galaxy, almost face-on, with spiral arms complicating the 
lens analysis (York et al 2005).
Due to extinction, the distance between
the two images A and B, is 317 mas in optical, while 335 mas
 in radio. Monitoring the time-delay, together
with a lensing model, taking into account the
spiral arms, leads to an estimation of the Hubble constant
of H$-0$ = 70 km/s/Mpc
(while 61 km/s/Mpc if spiral arms are masked out).

{\bf PKS1830-211}

Towards PKS1830-211, the lensing galaxy at z=0.88582 splits
the background source in two images A and B, each absorbed by 
a different velocity component (Frye et al 1997,
Wiklind \& Combes 1998). The intrinsic temporal
variability allows to monitor the time delay
between the two components.

Even without resolving spatially the two images,
it is possible to follow the intensity ratio between 
the two, since they are absorbing at two different velocities.
The IRAM monitoring during 3 years (1h per week) led to
a time delay of 24$\pm$5 days, and and estimation of H$_0$ 
= 69 $\pm$12 km/s/Mpc (Wiklind \& Combes 1999).

\begin{figure}
\centering
\includegraphics[width=4.5cm]{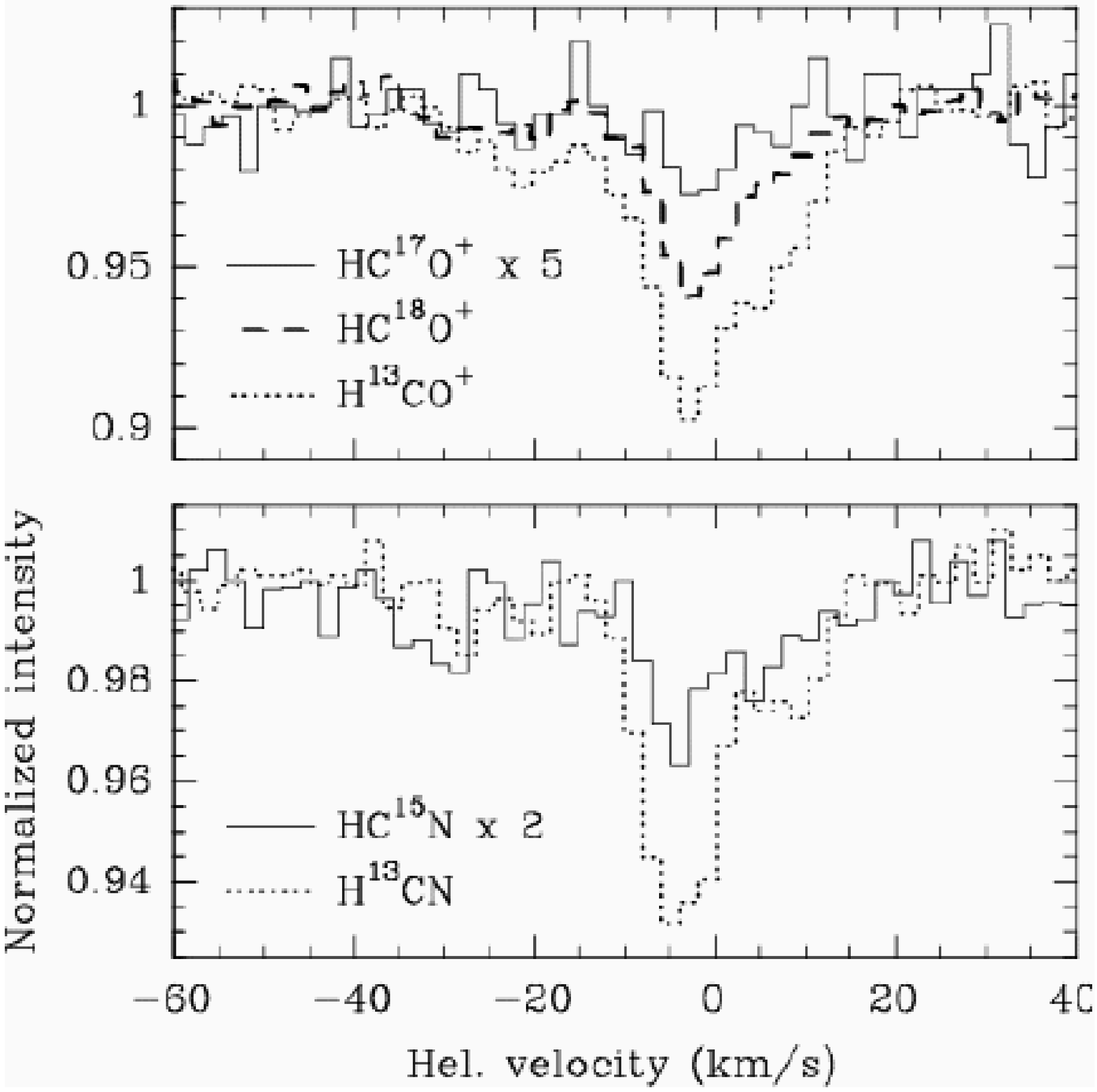}
\includegraphics[width=6cm]{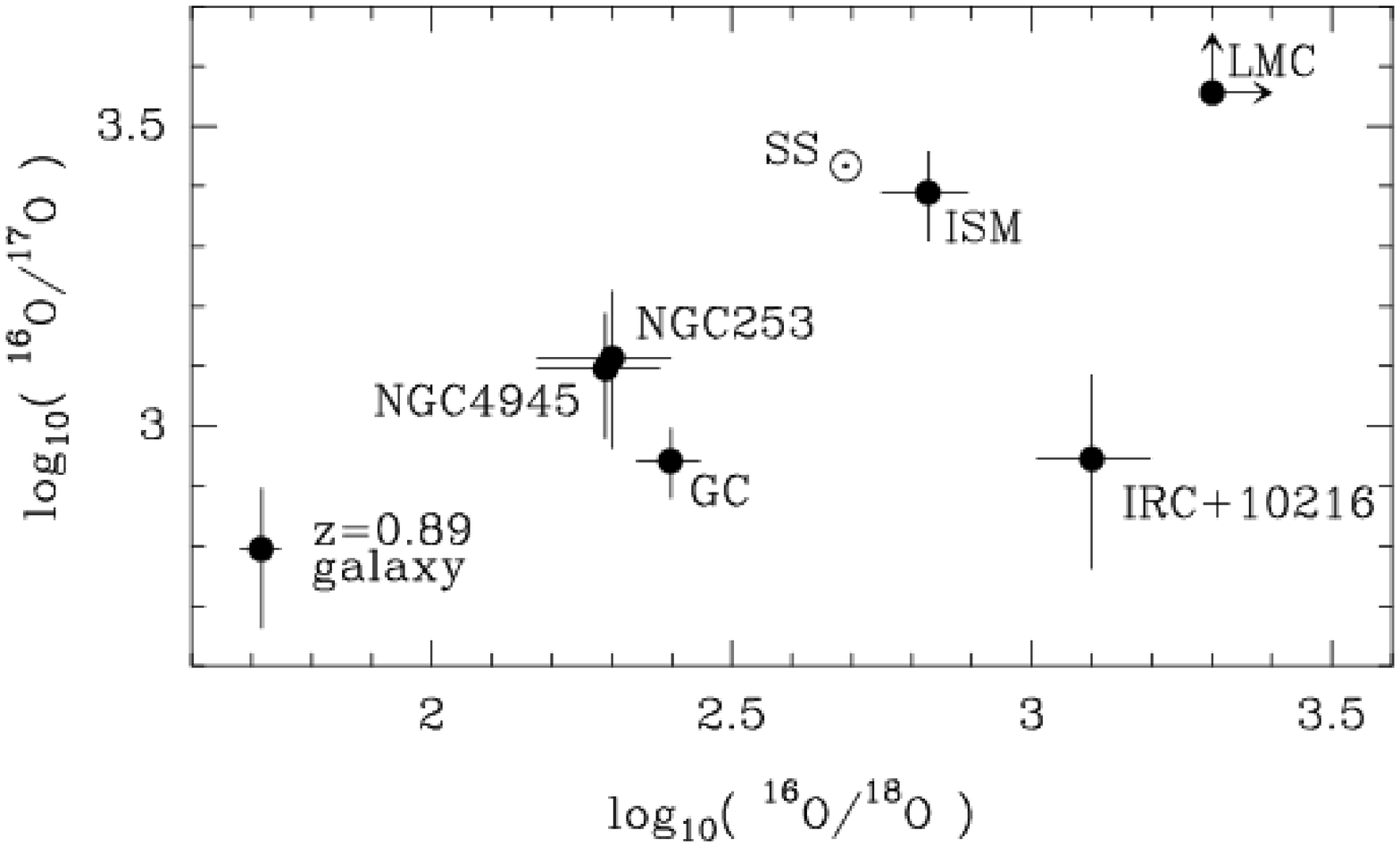}
\caption{{\bf Left} Molecular absorption lines including isotopes, like
$^{15}$N, $^{17}$O or $^{18}$O, in PKS1830-211 by M\"uller et al (2006).
{\bf Right} Oxygen isotopic ratios $^{16}$O/$^{17}$O versus  $^{16}$O/$^{18}$O for different sources.
The relative abundances of the oxygen isotopes at z=0.89 suggests that enrichment by low mass stars
had not yet time to dominate in this young lens galaxy, in front of PKS1830-211.}
\label{fig:2}
\end{figure}

The study of a large variety of
molecules allows to tackle the evolution of chemical conditions.
There does not seem to be variations at high z  in comparison with 
z=0 absorptions, but there is a large scatter, 
even locally (Lucas \& Liszt 1994, Liszt et al. 2006).

Upper limits were reported for deuterated molecules
(Shah et al 1999) and for CI (Gerin et al 1997).
A recent survey with the IRAM interferometer of
several isotopes (C, N, O or S) begins to find evidence
for abundance evolution (M\"uller et al 2006, Figure 2).

Only low excitation diffuse gas is observed on the line of sight
of PKS1830-211, the volumic density is so low that 
 T$_{ex} \sim$ T$_{CMB}$. The observation of several lines
of the rotational ladder of the same molecule (HCN, HNC, N2H$^+$,
H$^{13}$CO$^+$, CS..) can then lead to a measure of  T$_{CMB}$.
 Millimeter  absorptions can then complement the measurement
of T$_{CMB}$(z)  obtained from UV H$_2$  lines
(Srianand et al 2000, Reimers et al 2003, Cui et al 2005).

{\bf PMN J0134-0931}

This recent absorber has been detected in HI and OH lines
at GBT (Kanekar et al 2005),
from the  z=0.7645 lens in front of the background
quasar at z=2.22. The latter is split in 6 apparent images.
 Surprisingly, only upper limits of HCO$^+$ or H$_2$CO lines
were obtained on this source, probably due to small-scale structure of
the ISM, and very different continuum source extent across the radio spectrum.
  The absorption system provides a good probe of the fundamental
constant variation.

\vskip  -6mm
\subsection{Variations of constants}

Although laboratory measurements and 
solar system observations (e.g. Olive et al 2002,
Uzan 2003) do not show evidence for time variations
of the fundamental constant $\alpha$, high-z observations
have revealed a possible variation over larger time-scales
and also with space (Webb et al 2001, Murphy et al 2003)
with some controversy (Chand et al 2006, Tzanavaris et al 2006).
While from Alkali Doublet (CIV, SiII, SiIV, MgII, AlIII,..) on 22 absorbing systems 
and the ``many-multiplet'' method on 143 systems, a positive result
$\Delta\alpha/\alpha$ = (-0.6$\pm$0.1) 10$^{-5}$
has been claimed (Murphy et al 2003), this has not been confirmed
with the same method, $\Delta\alpha/\alpha$ = (-0.05$\pm$0.2) 10$^{-5}$
by Chand et al (2006). Independent methods with radio
lines are then welcome to better understand the
systematics of the various techniques.

The radio domain has the big advantage of
heterodyne techniques, with a  spectral resolution of 10$^6$ or more,
and dealing with cold gas and narrow lines.
Also different constants can be probed, while comparing the optical lines
with the HI 21cm , the OH 18cm and CO or HCO$^+$ rotational lines,
which depend very differently on $\alpha$, the electron-proton mass
ratio $\mu$ =m$_e$/m$_p$, or the proton gyromagnetic ratio g$_p$.
 In PKS1413+135, a resolution of 40m/s is required to resolve the lines,
and the obtained upper limits for variations on y = $\alpha^2$ g$_p \mu$ are
$\Delta$y/y = (-0.20 $\pm$ 0.20)  10$^{-5}$  and
$\Delta$y/y = (-0.16 $\pm$0.36)  10$^{-5}$  for B0218+357 (Murphy et al 2001).
The main systematics is the kinematical bias, i.e. that the different
lines do not come exactly from the same material along the line of sight, with
the same velocity.  Statistics with
absorptions of HI and HCO$^+$ in our own Galaxy in front of
remote quasars  (Lucas \& Liszt 1998) have measured a
dispersion of about 1.2km/s, corresponding to $\Delta$y/y = 0.4  10$^{-5}$.
 The results combining lines in  
PMN J0134-0931 and B0218+357 on  F = g$_p [\alpha^2/\mu]^{1.57}$ are 
$\Delta$F/F = (0.44 $\pm$0.36$^{stat} \pm$1.0$^{syst}$) 10$^{-5}$  for 
0 $<$ z  $<$ 0.7 (with statistical and systematical errors separated).
No  variation is detected, while the
sensitivity at 2$\sigma$  on the $\alpha$ variation is
 $\Delta\alpha$/$\alpha$ $\sim$6.7 10$^{-6}$, and on the mass ratio
 $\Delta\mu/\mu \sim$1.4 10$^{-5}$ over half of the age of the universe
(Kanekar et al 2005). It is then needed to find much more sources with ALMA.

\vskip  -6mm
\section{Perspectives}
\label{sec:3}

The number of molecular absorptions so far
(5 at high-z) and also the number of HI-21cm
absorbers (about 50 for z$>$0.1) is surprisingly low.
Why so few radio absorbers?
One explanation, at least for the molecular absorptions,
is that the high column density expected obscures the background
quasars, introducing a strong bias against the optical detection of these
remote sources. At least some could be known only in radio, but with no
redshift available. Curran et al (2006) have noticed a strong correlation
between the molecular fraction and the red colors of the quasars.
Those with molecular absorptions are in general compact 
flat spectrum sources, where most of the emission is covered.
Future searches should concentrate on sub-DLA systems, 
where the H$_2$ fraction  is higher, as well as
 metals  (Khare et al 2006, Kulkarni et al 2006). 

\begin{figure}
\centering
\includegraphics[width=5.5cm]{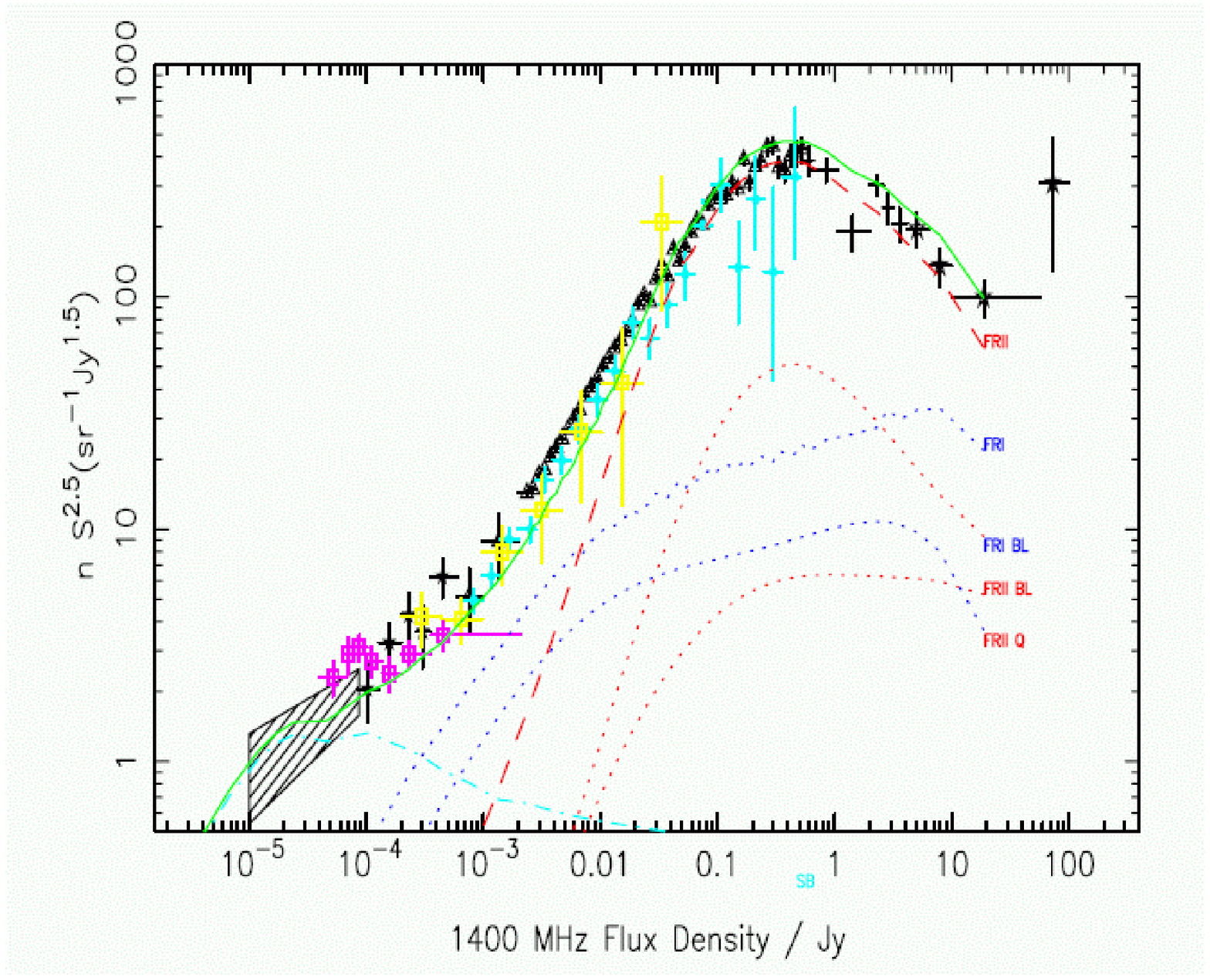}
\includegraphics[width=5.5cm]{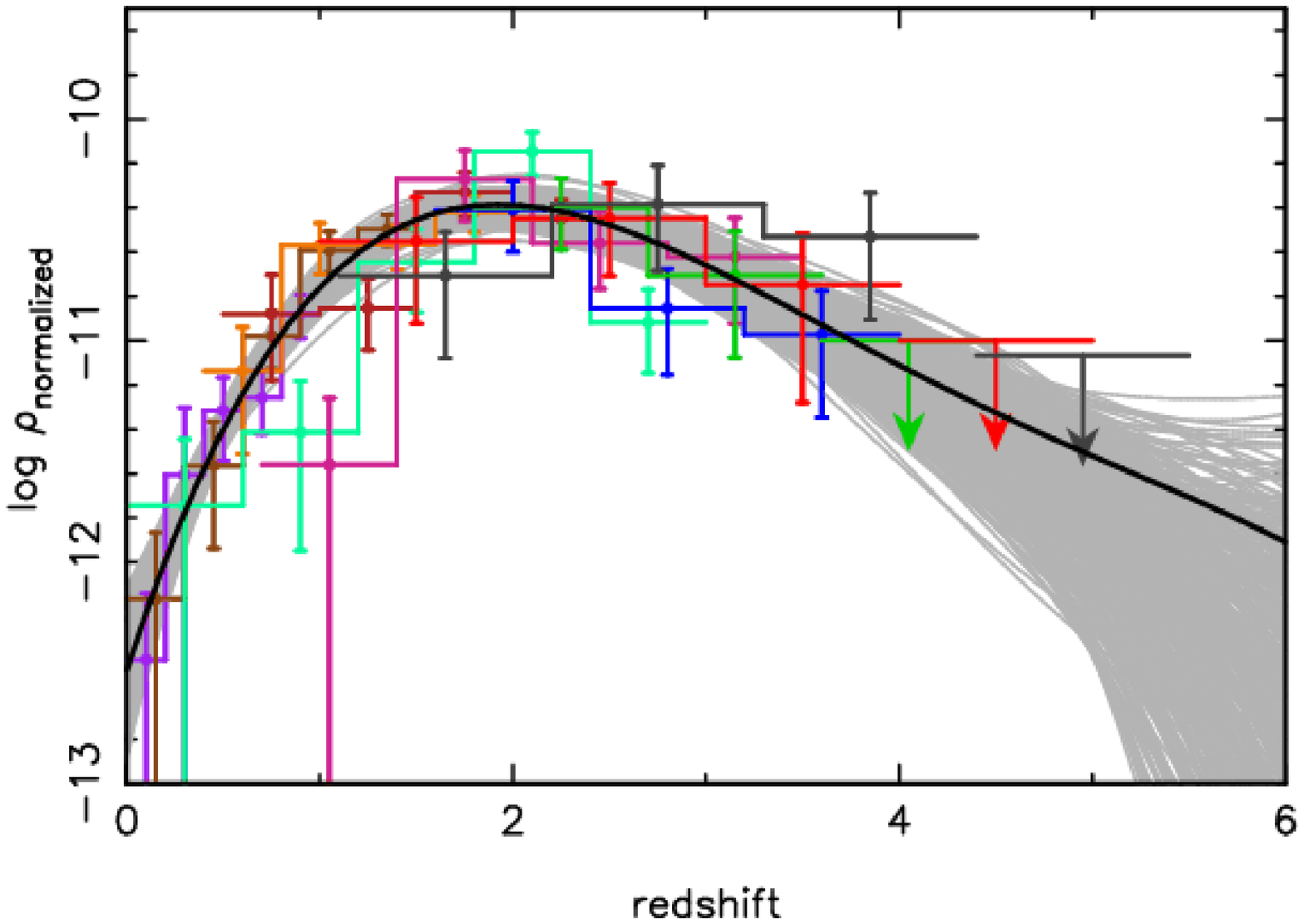}
\caption{The distribution of radio continuum sources,
from several observed samples (as the Parkes flat-spectrum sample), 
fitted with models
as a function of flux ({\bf Left}, Jackson 2004), and 
redshift ({\bf Right}, Wall et al 2005).
There is a strong increase in the density of radio-sources
until z=2, which translates into a strong increase in density of sources
with decreasing flux, above what is expected for the euclidean count
(falloff dn/dS as S$^{-5/2}$).}
\label{fig:3}
\end{figure}

Typical first projects with ALMA could be (included in the DRSP):
{\bf 1--} Molecular survey of PKS1413, PKS1830, CenA, in
   7 wide priority bands, with spectral resolution of 1-4 km/s; 
{\bf 2--} Search for new systems,
towards 60 selected radio loud AGNs with mm cont flux $>$ 50mJy,
with criteria of obscuration, gravitational lensing and/or
suppressed soft X-ray flux. When no redshift is known, the
search could be over the entire redshift range using the technique
of frequency scanning.

As shown in Figure 3, it is now well-known that the volumic density
of radio quasars peaked around z=2 (Shaver et al,  1996, Wall et al 2005),
and there is a cutoff after z=3.
Optical quasars follow the same curve, in a
similar way to the star formation history.
 In parallel, the number of sources as a function of flux
N(S) increases well above the euclidean curve in S$^{-1.5}$,
and we could expect to detect 1 or 2 orders of magnitude
more quasars with ALMA.
However at high-z, their millimeter flux is weakened by the 
non-favorable K-correction (compact and flat-spectrum sources being
rare). In this domain, it is  interesting to search 3mm systems at cm wavelengths,
with Band 1 and 2 of ALMA in the future.

%

\end{document}